\documentclass[twocolumn,aps,superscriptaddress,showpacs,prl]{revtex4}
\usepackage{graphicx}
\usepackage{times}
\usepackage{amsmath}
\usepackage{amsfonts}
\usepackage{amssymb}
\usepackage{units}
\usepackage{stmaryrd} 
\DeclareGraphicsExtensions{.pdf,.png,.eps,.jpg}

\begin{document}
\preprint{0}

\title{Giant Ambipolar Rashba Effect in a Semiconductor: BiTeI}

\author{A. Crepaldi}
email[alberto.crepaldi@epfl.ch]  \affiliation{Institute of Condensed Matter Physics (ICMP), Ecole Polytechnique
F\'ed\'erale de Lausanne (EPFL), CH-1015 Lausanne,
Switzerland}

\author{L. Moreschini}\affiliation{Advanced Light Source (ALS), Lawrence Berkeley National Laboratory, Berkeley, California 94720, USA}

\author{G. Aut\`es}
\affiliation{Institute of Theoretical Physics, Ecole Polytechnique F\'ed\'erale de Lausanne (EPFL), CH-1015 Lausanne,
Switzerland}

\author{C. Tournier-Colletta}\affiliation{Institute of Condensed Matter Physics (ICMP), Ecole Polytechnique
F\'ed\'erale de Lausanne (EPFL), CH-1015 Lausanne,
Switzerland}

\author{S. Moser}
\affiliation{Institute of Condensed Matter Physics (ICMP), Ecole Polytechnique
F\'ed\'erale de Lausanne (EPFL), CH-1015 Lausanne,
Switzerland}

\author{N. Virk}
\affiliation{Institute of Theoretical Physics, Ecole Polytechnique
F\'ed\'erale de Lausanne (EPFL), CH-1015 Lausanne,
Switzerland}

\author{H. Berger}
\affiliation{Institute of Condensed Matter Physics (ICMP), Ecole Polytechnique
F\'ed\'erale de Lausanne (EPFL), CH-1015 Lausanne,
Switzerland}

\author{Ph. Bugnon}
\affiliation{Institute of Condensed Matter Physics (ICMP), Ecole Polytechnique
F\'ed\'erale de Lausanne (EPFL), CH-1015 Lausanne,
Switzerland}

\author{Y. J. Chang}\affiliation{Advanced Light Source (ALS),  Lawrence Berkeley National Laboratory, Berkeley, California 94720, USA}

\author{K. Kern}\affiliation{Institute of Condensed Matter Physics (ICMP), Ecole Polytechnique
F\'ed\'erale de Lausanne (EPFL), CH-1015 Lausanne,
Switzerland}
\affiliation{Max-Plank-Institut f\"ur Festk\"orperforschung,
D-70569, Stuttgart, Germany}

\author{A. Bostwick}\affiliation{Advanced Light Source (ALS), Lawrence Berkeley National Laboratory, Berkeley, California 94720, USA}

\author{E. Rotenberg}\affiliation{Advanced Light Source (ALS), Lawrence Berkeley National Laboratory, Berkeley, California 94720, USA}

\author{O. V. Yazyev}
\affiliation{Institute of Theoretical Physics, Ecole Polytechnique
F\'ed\'erale de Lausanne (EPFL), CH-1015 Lausanne,
Switzerland}

\author{M. Grioni}
\affiliation{Institute of Condensed Matter Physics (ICMP), Ecole Polytechnique
F\'ed\'erale de Lausanne (EPFL), CH-1015 Lausanne,
Switzerland}
\date{\today}

\begin{abstract}
We observe a giant spin-orbit splitting in bulk and surface states of the non-centrosymmetric semiconductor BiTeI. We show that the Fermi level can be placed in the valence or in the conduction band by controlling the surface termination. In both cases it intersects spin-polarized bands, in the corresponding surface depletion and accumulation layers. The momentum splitting of these bands is not affected by adsorbate-induced changes in the surface potential. These findings demonstrate that two properties crucial for enabling semiconductor-based spin electronics -- a large, robust spin splitting and ambipolar conduction -- are present in this material.
\end{abstract}

\pacs{73.20.At, 71.70.Ej,79.60.Bm, 79.60.-i}

\maketitle

The relativistic spin-orbit interaction (SOI) lifts the usual Kramers spin degeneracy in electron systems that lack inversion symmetry. It lies at the origin of many subtle and interesting effects in the electronic structure of 
materials such as the emergence of topological insulators (TI), a new quantum state of matter. In the bulk of materials with non-centrosymmetric structures, such as the zincblend and wurzite structures, it gives rise to the Dresselhaus \cite{Dress_PR_1955} and Rashba \cite{Rashba_1960} effects. An analogous effect, the Rashba-Bychkov effect, describes the lifting of the spin degeneracy at surfaces and at asymmetric interfaces, where inversion symmetry is also broken \cite{Rashba_JETP_1998}. The SOI is a general phenomenon, but it is especially relevant  in solids containing high-$\textrm Z$ elements because of their large atomic spin-orbit parameter. The characteristic splitting in energy and momentum was first directly observed by angle-resolved photoelectron spectroscopy (ARPES) on the Au(111) surface  \cite{lashell_PRL_1996}. The predicted polarization of the electronic states was confirmed by spin-polarized ARPES \cite{Hochstrasser_PRL_2002,Hoesch_PRB_2004}, and the Rashba scenario has been extended to other surfaces and interfaces \cite{rot_PRL_1999,Koroteev_PRL_2004,Krupin_PRB_2005,Cercellier_PRB_2006,Barke_PRL_2006,ast_PRL_2007,He_PRL_2008,
Sakamoto_PRL_2009,gierz_PRL_2009}.

The vision of an all-electric control of spin transport in new device concepts explains the strong current interest for materials with large Rashba or Dresselhaus effects. 
Future devices operating at room temperature will require a large separation between the spin-polarized bands and the ability to tune the position of the chemical potential over a broad energy range. Whereas the former have been reported in surface alloys with
high-$\textrm Z$ elements such as Pb or Bi, only limited tunability has been achieved so far. 

BiTeI is a non-centrosymmetric semiconductor for which theory predicts a large bulk Rashba effect, and the emergence of a topological insulating phase under pressure \cite{Bahramy_NatureComm_2012}. Ishizaka {\it et al.} \cite{ishizaka_NAT_2011} used spin-resolved ARPES to reveal spin-polarized states with a large momentum splitting. They assigned them to a quantum-well state (QWS) confined in the accumulation layer that appears because of band bending in the surface region. This interpretation has been questioned in part by more recent ARPES data and theory that show the coexistence of surface and bulk bands near the Fermi level \cite{Landolt_arXiv_2012}.

In this Letter, we show that in BiTeI the chemical potential can be moved well into the conduction band or the valence band  by controlling the surface termination. Remarkably, a giant spin splitting at the Fermi surface is observed in both cases. First-principles relativistic calculations indicate that both the surface and the bulk bands are split by the SOI. We also prove that size of the Rashba effect is largely insensitive to changes in the surface potential. Therefore, the splitting has mainly an atomic origin.  These results establish BiTeI as a versatile material, characterized by the coexistence of very large ambipolar bulk and surface Rashba effects. 

\begin{figure*}[!t]
  \centering
  \includegraphics[width =16 cm]{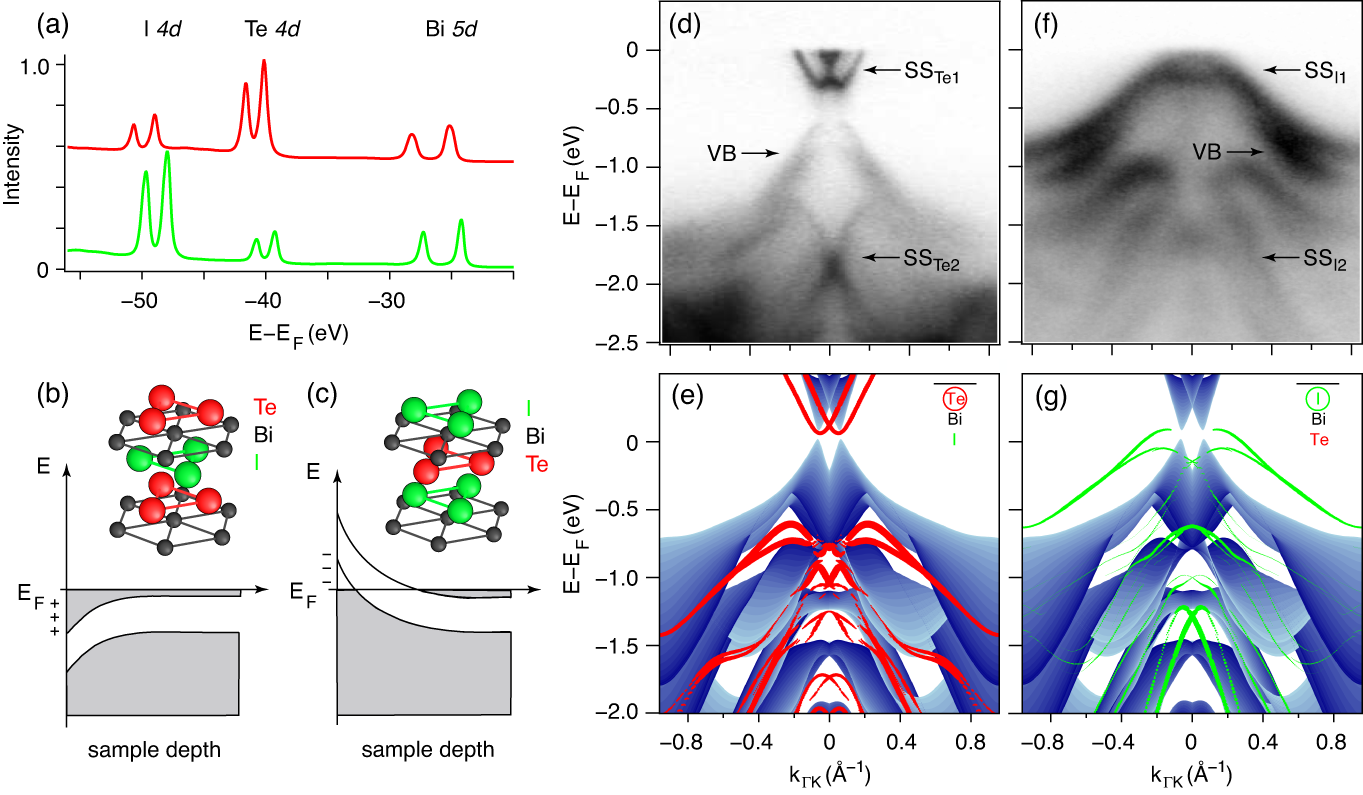}
  \caption[didascalia]{(Color online) (a) I $4d$, Te $4d$ and Bi $5d$ core level spectra measured at h$\nu =120$ eV on Te- (red, top) and I-terminated (green, bottom) surfaces. (b,c) Schemes of surface band bending for the two surface terminations.  (d) ARPES intensity map along the $\overline{\Gamma \textrm K}$ direction measured at $93$ eV and $\textrm T=40~\textrm K$, compared to (e) the projected slab band structure calculated from first principles. 
The size of red symbols is proportional to the magnitude of projection onto the surface Te atoms indicated in the inset. The continuum of bulk states is shown in blue.
(f,g) Corresponding plots for the I-terminated surface. The size of the green symbols is proportional to the contribution of the surface I atoms indicated in the inset.
}
  \label{Fig1}
\end{figure*}

We performed ARPES experiments at the Electronic Structure Factory, beam line 7.0.1 of the Advanced Light Source. 
The energy and momentum resolution of the hemispherical Scienta R4000 analyzer were 30 meV and 0.1$^{\circ}$. High quality single crystals of BiTeI, in the form of platelets, were grown by chemical vapor transport and by the Bridgman technique, and characterized by x-ray diffraction and transport. They showed a metallic conduction due to a small ($<$2\%) deviation from stoichiometry. The samples were mounted on a He cryostat and cleaved in UHV to expose flat, shiny surfaces.

First-principles electronic structure calculations were performed within 
the density functional theory (DFT) framework employing the generalized gradient
approximation (GGA) as implemented in the {\sc Quantum-ESPRESSO} package \cite{QE}. Spin-orbit effects were accounted for using the fully relativistic norm-conserving pseudopotentials acting on valence electron wavefunctions represented in the two-component spinor form \cite{DalCorso05}.
The surface band structures were obtained using a slab model consisting of 39 atomic layers.
Since the BiTeI crystal has no inversion symmetry, the surfaces of the slab are necessarily different. The model considered includes two experimentally relevant terminations (Te and I). The surface bands were extracted from the slab band structure by projecting the Kohn-Sham wave functions onto atomic wave functions at the surface layer. 
The bulk and the slab band structures were aligned by matching the potential in the middle of the slab with the bulk potential.


BiTeI has a trigonal layered structure, with Bi, Te and I planes alternating along the $c$ axis. The Bi and Te planes are covalently bonded to form a positively charged (BiTe)$^+$ bilayer. The ionic coupling between the bilayer and the adjacent I$^{-}$ plane defines the natural cleavage plane \cite{sheve_XRAY_1995}. The topmost layer -- Te or I -- is identified by the relative intensities of the Te and I $4d$ core levels, as in Fig. 1(a). Ideally, due to the lack of inversion symmetry, the surface termination  is uniquely determined by the direction of the $c$-axis.  However, repeated cleaves of the same crystal randomly expose both terminations due to the occurrence of stacking faults, which also explains the observation of 6-fold symmetry in the Laue patterns (not shown). We have measured several `pure' surfaces and on `mixed' surfaces that presented areas with both terminations \cite{note}. Data for the former are illustrated in Fig. 1.
The surface charge -- positive for Te, negative for I -- induces band bending in opposite directions for the two terminations.  The Fermi level lies into either the conduction or the valence band, giving rise to a charge accumulation or, respectively, depletion layer. This is schematically illustrated in Fig.~1(b,c) and substantiated by the ARPES data.

Figure 1(d) is an ARPES intensity map of the Te-terminated surface, measured along the $\overline{\Gamma \textrm K}$ high-symmetry direction ($\overline{\Gamma \textrm K} = 0.82$~\AA$^{-1}$) of the Brillouin zone. The most prominent feature is the split parabolic band (SS$_{\textrm Te1}$) straddling the Fermi level E$_{\textrm F}$. The two subbands have minima at $-0.32$ eV and are offset by $\pm 0.05~\textrm \AA^{-1}$ around  $\overline{\Gamma}$. This is consistent with previous data, and with a Rashba interaction one order of magnitude stronger than for the Au(111) benchmark case \cite{ishizaka_NAT_2011}. This feature is well reproduced in the first-principles band structure of Fig.~1(e). It shows that the spin-split state is localized in the topmost bilayer, and partially overlaps with conduction band states that exhibit a smaller momentum offset. 
The bulk signal is too weak to be identified in Fig. 1(d), but it can be discerned between SS$_{\textrm Te1}$ and E$_{\textrm F}$ at closer inspection \cite{note}. The projected bulk valence band exhibits gaps supporting other surface states. These states exhibit an even larger, but previously unnoticed, splitting.  Their spin polarization has non negligible radial and out-of-plane components \cite{note}, at variance with the simple Rashba scenario, and similar to recent observations on topological insulators \cite{Xia_Science_2011}. Only part of this manifold, labeled SS$_{\textrm Te2}$, is visible in the ARPES map, due to matrix elements and to the overlap with the bulk continuum. One can readily recognize symmetrically split subbands, with maxima at $\pm 0.2$~\AA$^{-1}$
and  $-1.3$~eV, crossing at $\overline{\Gamma}$ at $-$1.8~eV. Bulk valence band states are visible between SS$_{\textrm Te1}$ and SS$_{\textrm Te2}$, with a maximum $0.37$~eV below the bottom of SS$_{\textrm Te1}$. These states are also offset in momentum, reflecting a large bulk spin-orbit splitting.

The picture from the I-terminated surface (Fig.~1(f,g)) is quite different. The electron pockets around $\overline{\Gamma}$ are replaced by hole pockets from a spin-split state (SS$_{\textrm I1}$) with a strong projection on the surface I atoms. The momentum offest is again quite large, of the order of  $\pm 0.2$~\AA$^{-1}$. A precise determination is difficult because the top of the band lies above E$_{\rm F}$. The top of VB is also located above E$_{\rm F}$. There is a complete change from electron to hole carriers with respect to Fig.~1(d), which demonstrates ambipolar conduction in BiTeI. The total change in band bending between the Te- and I-terminated surfaces, estimated from core level spectra \cite{note}, is $\Delta \textrm E_{\rm BB}=0.9$~eV, to be compared with the estimated energy gap
$\Delta \textrm E_\textrm g\sim0.38$ eV \cite{ishizaka_NAT_2011}. Here again, additional surface features, partially overlapping the valence band continuum, can be identified at higher energy both in the ARPES map (SS$_{\textrm I2}$) and in the calculations.

\begin{figure}[!t]
  \centering
  \includegraphics[width = 8 cm]{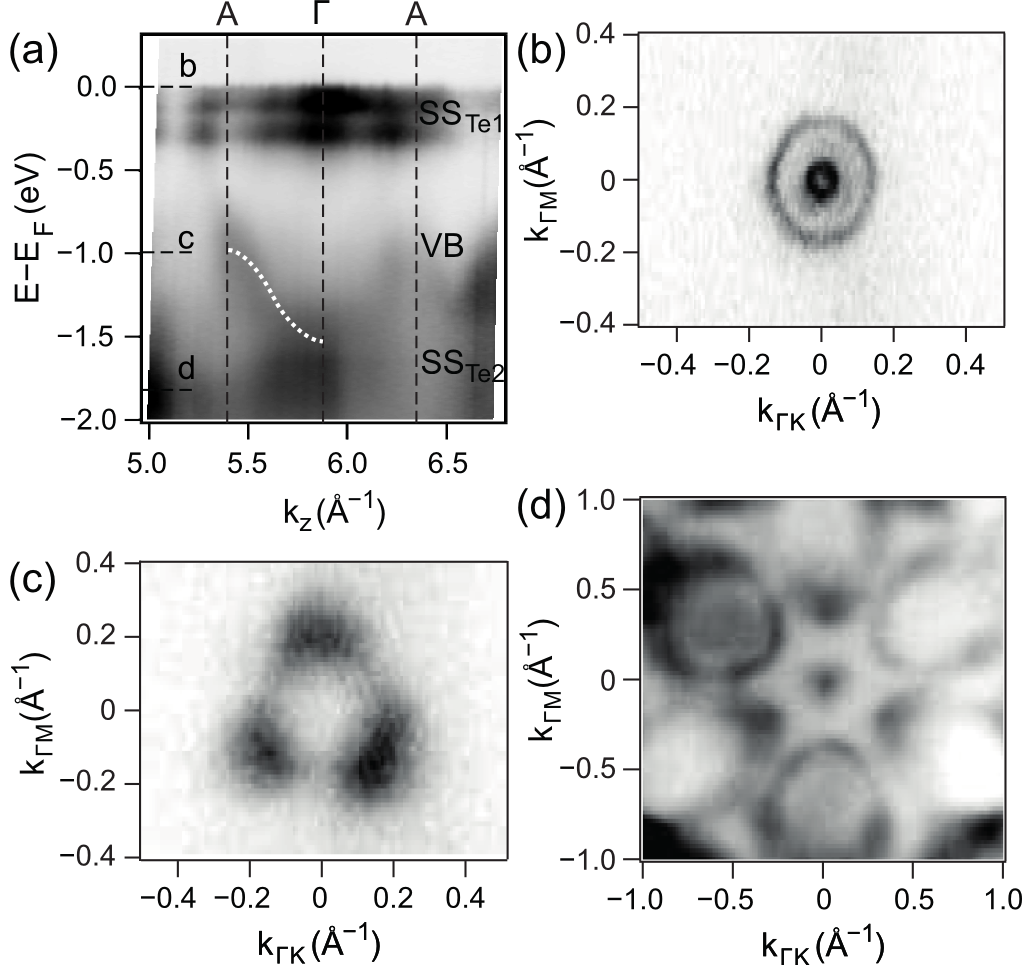}
  \caption[didascalia]{(Color online) (a) ARPES intensity from a Te-terminated surface, measured at $\sim 0.3^\circ$ off normal emission,  in a photon energy scan between $84$ eV and $162$ eV, plotted as a function of $k_\perp$, the wave vector along the $c$-axis. Both the upper and the lower branch of SS$_{\textrm Te1}$ are visible. The white dashed curve is the calculated dispersion of the corresponding bulk valence state. (b-d) Constant energy contours measured at the energies marked by the corresponding (b,c,d) horizontal lines in panel (a).}
  \label{fig:arpes2}
\end{figure}

The assignment of the spectral features of Fig.~1 to surface or bulk states is further supported by the data shown in Fig.~2(a). It illustrates the photoelectron intensity measured at near-normal emission from a Te-terminated surface as a function of $k_\perp$, the wave vector along the $c$-axis. Bands SS$_{\textrm Te1}$ and SS$_{\textrm Te2}$ exhibit an intensity modulation but no dispersion, as expected for true surface states. By contrast, state VB exhibits a $\sim 0.7$ eV dispersion along $\Gamma \textrm A$, consistent with its bulk character, and well reproduced by the calculation (dashed line). This is confirmed by
constant energy maps (CEM) of these states. The CEM measured at E$_\textrm F$ for SS$_{\textrm Te1}$ (Fig.~2(b)) has two concentric contours typical of the Rashba scenario. The external contour, warped by the interaction with the lattice potential, has a 6-fold symmetry, consistent with time-reversal symmetry for a surface state \cite{Fu_PRL_2009}. In the CEM measured at the crossing point of SS$_{\textrm Te2}$ (Fig 2(d)), the inner contour has collapsed to a point at $\overline{\Gamma}$. The outer contour again exhibits a 6-fold symmetry. Interestingly, it is not closed around $\overline{\Gamma}$, but it is broken into 6 disconnected pockets aligned along the 6 equivalent $\overline{\Gamma \textrm M}$ directions. This can be seen as the limit of strong warping, reflecting a large in-plane asymmetry of the surface potential \cite{Emm_PRB_2011}. Conversely, the CEM through VB (Fig. 2(c)) shows a single 3-fold contour. This symmetry reduction is not due to partial extinction of a 6-fold contour induced by ARPES matrix-elements, because the pattern remained locked to the crystallographic directions when the crystal was rotated around the surface normal. Therefore, the 3-fold pattern of VB reflects the 3-fold symmetry of the bulk potential, and confirms the bulk character of this state.

We now turn our attention to the origin of the very large Rashba splitting. Competing models beyond the standard Rashba scenario have been proposed. They alternatively stress atomic contributions \cite{Petersen_SS_2000}, the in-plane anisotropy of the surface potential \cite{henk_PRB_2007}, the asymmetry of the wavefunctions \cite{Bihlmayer_PRB_2007}, or the local orbital angular momentum \cite{Park_PRL_2011}. A bulk origin has been invoked for BiTeI \cite{ishizaka_NAT_2011}, but this conflicts with the surface nature of the relevant states. 

In the standard Rashba scenario the size of the splitting is controlled by the gradient of the surface potential,
and this prediction was found to be consistent with the properties of QWS formed in an accumulation layer at the surface of the TI Bi$_2$Se$_3$ \cite{king_PRL_2011}.
In order to test this hypothesis for BiTeI we have changed in a controlled way the surface band bending, and hence the gradient of the potential in the surface region. This was achieved by depositing increasing amounts of potassium on a Te-terminated surface.
Adsorbed K atoms donate electrons to the CB, leaving a positively charged surface layer which enhances the downward surface band bending. Figure 3(a) illustrates the evolution of SS$_{\textrm Te1}$ as a function of K coverage. Movies of complete K dosing experiments for both Te- and I-terminated surfaces are available in Ref. \onlinecite{note}. 

As expected, SS$_{\textrm Te1}$, as well as all core levels \cite{note}, shift to lower energies, following the change in band bending. The total shift at saturation K coverage is 0.12~eV, bringing the bottom of SS$_{\textrm Te1}$ band 0.44~eV below E$_\textrm F$. 
A closer inspection shows that the energy shift of SS$_{\textrm Te1}$ is rigid. Figure~3(b) shows the Fermi wave vector $k_\textrm F$ of the outer branch and the momentum offset $k_0$. They were estimated  from a parabolic fit of the dispersion, keeping the effective mass unchanged. Within error bars $k_0$ remains constant. The bottom of SS$_{\textrm Te1}$ gives an upper limit for band bending.
Therefore a $> $40\% change in the surface band bending has no measurable effect on the strength of the Rashba effect. This
experimental observation 
strongly suggests that other parameters, namely the atomic spin-orbit parameter of the heavy elements, determine the large spin splitting.
Figure 3(c) shows that the surface carrier density, estimated from the area of the electron pockets, varies linearly with the downward shift of the split bands. This is again consistent with a constant $k_0$. By contrast, deviations from linearity have been observed for the QWS at the Bi$_2$Se$_3$ surface \cite{king_PRL_2011}.

\begin{figure}[t]
  \centering
  \includegraphics[width = 8 cm]{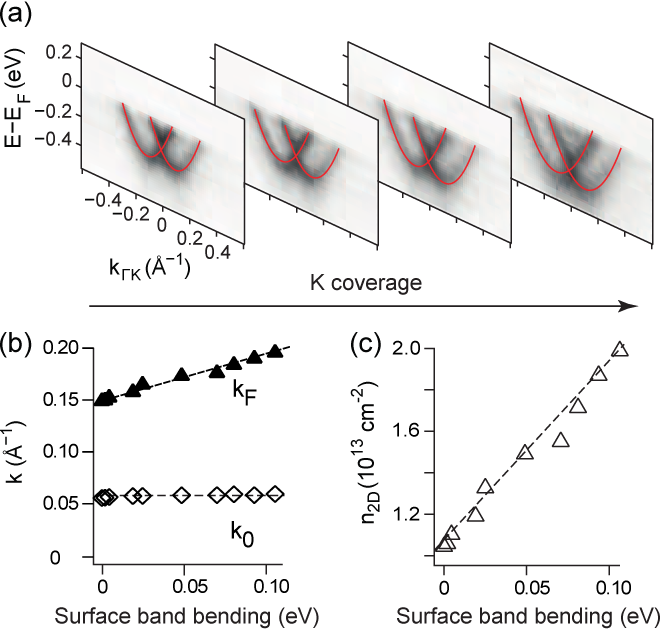}
  \caption[didascalia]{(Color online) (a) Evolution of the spin-split SS$_{\textrm Te1}$ surface state band as a function of K coverage. The red parabolas are the result of a fit to the band dispersion along $\overline{\Gamma \textrm K}$. (b) The Fermi wave vector $k_\textrm F$ of the outer branch (solid symbols) and the Rashba splitting $k_0$ (empty symbols) are plotted as a function of the measured surface band bending.
(c) The corresponding surface electron density.}
  \label{fig:arpes3}
\end{figure}

In summary, we have shown that large ambipolar bulk and surface Rashba effects coexist in the non-centrosymmetric semiconductor BiTeI. The Fermi level at the surface lies either into the valence or the conduction band, depending on the nature of the topmost layer. The surface termination
is randomly chosen by cleavage due to stacking faults in bulk crystals, but it could be precisely determined in artificially grown
thin films. It is realistic to consider that regions with opposite band bending could then be patterned on a substrate, opening new perspectives for the manipulation of spin-polarized states.

We acknowledge support by the Swiss NSF, namely through grants No. PP00P2\_133552 (G.A., N.V. and O.V.Y) and PBELP2-125484 (L.M.). The calculations were performed at the CSCS (Manno). The Advanced Light Source is supported by the Director, Office of Science, Office of Basic Energy Sciences, of the U.S. Department of Energy under Contract No. DE-AC02-05CH11231.


\end{document}